\documentclass[11pt]{article}
\usepackage{graphicx,caption,amsmath}

\newcommand{\BABARPubYear}    {01}

\newcommand{\BABARConfNumber} {29}
\newcommand{\SLACPubNumber} {8983}

\input pubboard/babarsym

\long\def\inst#1{\par\nobreak\kern 4pt\nobreak
    {\it #1}\par\vskip 10pt plus 3pt minus 3pt}

\begin{document}
\pagestyle{empty}

\begin{flushright}
\babar-CONF-\BABARPubYear/\BABARConfNumber \\
SLAC-PUB-\SLACPubNumber\\
September, 2001
\end{flushright}

\par\vskip 4.75cm

\begin{center}
\Large \bf Search for a Lifetime Difference in $D^0$ Decays
\end{center}
\bigskip

\begin{center}
\large The \lbabar\ Collaboration\\
\mbox{ }\\
\today
\end{center}
\bigskip \bigskip

\begin{center}
\large \bf Abstract
\end{center}
The $D^0$ mixing parameter
$y = \Delta \Gamma/2 \Gamma$ was determined by measuring the $D^0$
lifetime separately for the $K^- \pi^+$ decay mode 
and the $K^- K^+$ decay
mode with 12.4\invfb of data collected
by the \babar\  experiment in 2001. Backgrounds were suppressed with the
$D^* \to D^0 \pi^+$ decay and particle identification.
The following preliminary result was obtained:
\begin{equation*} 
y = \left( -1.0 \pm 2.2 {\rm (stat)} \pm 1.7 {\rm (syst)}\right) \;\%\:.
\end{equation*}

\vfill
\begin{center}
Submitted to the 
9$^{th}$ International Symposium on Heavy Flavor Physics \\
9/10---9/13/2001, Pasadena, CA, USA
\end{center}

\vspace{1.0cm}
\begin{center}
{\em Stanford Linear Accelerator Center, Stanford University, 
Stanford, CA 94309} \\ \vspace{0.1cm}\hrule\vspace{0.1cm}
Work supported in part by Department of Energy contract DE-AC03-76SF00515.
\end{center}

\newpage
\pagestyle{plain}

\begin{center}
\small

The \babar\ Collaboration,
\bigskip

B.~Aubert,
D.~Boutigny,
J.-M.~Gaillard,
A.~Hicheur,
Y.~Karyotakis,
J.~P.~Lees,
P.~Robbe,
V.~Tisserand
\inst{Laboratoire de Physique des Particules, F-74941 Annecy-le-Vieux, France }
A.~Palano,
A.~Pompili
\inst{Universit\`a di Bari, Dipartimento di Fisica and INFN, I-70126 Bari, Italy }
% ...

\end{center}\newpage

\setcounter{footnote}{0}

\section{Introduction}

If \CP conservation holds in the $D^0$ system, the \CP-even and \CP-odd
eigenstates are mass eigenstates with widths $\Gamma_+$ and $\Gamma_-$,
respectively. The mixing parameter
$y = (\Gamma_+-\Gamma_-)/(\Gamma_++\Gamma_-)$ is a 
measure of the difference of
these lifetimes and is expected to be small ($10^{-3})$ within the Standard
Model~\cite{theory}.
If the observed value of $y$ is much larger than this expectation,
it could be difficult to reconcile with theory.
Otherwise, a small value of $y$ would be useful in constraining the size of
the other mixing parameter $x = 2(M_+-M_-)/(\Gamma_++\Gamma_-)$ in direct
measurements
of $D^0$ mixing, where $M_{\pm}$ are the masses of the \CP eigenstates.

A value of $y$ may be determined by measuring the lifetime for
$D^0$ mesons\footnote{In this paper, statements involving the $D^0$ meson
and its decay modes are intended to apply in addition to their
charged conjugates.}
that decay into final states of specific \CP
symmetry \cite{phenom}.
One such final state that is an equal mixture of \CP-even and \CP-odd 
is produced by the Cabibbo-favored decay $D^0 \rightarrow K^- \pi^+$.
If $y$
is small, the lifetime distribution of $D^0$ mesons decaying into
this final state can be approximated as an exponential with
lifetime $\tau_{K\pi} = 1/\Gamma$ where $\Gamma = (\Gamma_+ + \Gamma_-)/2$.

The $K^- \pi^+$ final state may be compared to $K^- K^+$ which is
\CP-even and is
produced by the Cabibbo-suppressed decay $D^0 \rightarrow K^- K^+$.
The lifetime distribution of $D^0$ mesons that decay
into $K^- K^+$ is exponential
with lifetime $\tau_{KK} = 1/\Gamma_+$. This lifetime can be compared
to $\tau_{K\pi}$ to obtain $y$:
\begin{equation}
y = \frac{\tau_{K\pi}}{\tau_{KK}} - 1 \;.
\label{eq:ycp}
\end{equation}
Since the $K^- \pi^+$ and $K^- K^+$ final states have similar topology,
many systematic uncertainties in the $D^0$ lifetimes cancel in their
ratio, making Eq.~\ref{eq:ycp} a particularly sensitive measurement.

Presented in this paper is a preliminary measurement of $y$ based 
on data collected with the \babar\  detector at the
\pep2 asymmetric $e^+e^-$ collider. The results were obtained from
a sample of 12.4\invfb
of 2001 data
that were reconstructed with the latest tracking alignment
parameters and reconstruction algorithms.
Data taken on and off the $\Upsilon(4S)$ resonance were used at a
center-of-mass that corresponds to a nominal Lorentz boost 
of $\beta \gamma = 0.56$ along the beam axis. 
The size of the interaction point (IP)
transverse to the beam direction was typically 6~$\mu$m in
the vertical direction and 120~$\mu$m in the horizontal direction.

\section{The \babar\ Detector}

The \babar\ detector,
a general purpose, solenoidal, magnetic spectrometer, is
described in more detail elsewhere \cite{babar}. Those detector components
employed in this analysis are briefly discussed here.
Charged particles were detected
and their momenta measured by a combination of a 40-layer drift chamber (DCH)
and a five-layer, double-sided,
silicon vertex tracker (SVT), both operating within a
1.5~T solenoidal magnetic field. $D^0$ decay vertices were typically
reconstructed with a resolution along the $D^0$ direction of 75~$\mu m$ for
two-prong decays. A ring-imaging Cherenkov detector (DIRC) was used for
charged particle identification.

\section{Event Selection}

The widths $\Gamma$ and $\Gamma_+$ were determined by fitting the decay time
distributions of independent samples of 
$D^0 \to K^- \pi^+$ and $D^0 \to K^- K^+$ decays.
The $D^0$ candidates of each sample were identified
from the charged particles belonging to their final state.
The decay $D^{*+} \rightarrow D^0 \pi^+$ and $K^\pm$ particle identification
were used to suppress backgrounds. 

$D^0$ candidates were selected by searching for pairs of tracks of
opposite charge and combined
invariant mass near the $D^0$ mass $m_D$. Each track
was required to contain a minimum number of SVT and DCH hits in order
to ensure their quality. The two $D^0$-candidate daughter
tracks were fit to
a common vertex. The $\chi^2$ probability of this vertex fit was
required to be better than 1\%.

Each $D^0$ daughter track that corresponds
to a $K^\pm$ particle was required to pass a likelihood-based particle
identification algorithm. This algorithm was based on the measurement
of the Cherenkov angle from the DIRC for momenta
$p \gsim 0.6$\gevc and on the energy loss ($dE/dx$) measured with
the SVT and DCH for momenta $p \lsim 0.6$\gevc.
The charged $K^\pm$ identification efficiency was approximately
80\% on average
for tracks within the DIRC acceptance with a $\pi^\pm$ misidentification
probability of about 2\%.

The decay $D^{*+} \rightarrow D^0 \pi^+$ is distinguished by a
$\pi^+$ of low momentum, commonly referred to as the slow pion
($\pi_s$). To increase acceptance, $\pi_s$ candidate tracks were not
required to contain DCH hits. To improve resolution, a vertex fit
was used to constrain each $\pi_s$
candidate track to pass through the intersection of the
$D^0$ trajectory and the IP. If the $\chi^2$ probability of this
vertex fit was less than 1\%, the $D^*$ candidate was discarded.

The $D^*$ candidates peak at a value of
$\delta m \approx 145.4$\mevcc, where $\delta m$ is the difference
in the reconstructed $D^*$ and $D^0$ masses. Backgrounds were rejected
by discarding events with a value of $\delta m$ that deviated more than 
a given margin from the peak. The size of this margin corresponded to
approximately three standard deviations and
varied between 1 and 2.5\mevcc depending on the quality of the $\pi_s$ track.

To remove background from $B$ meson decays, each $D^*$ candidate
was required to have a momentum $p^*$ in the center-of-mass
greater than 2.6\gevc. This requirement was also effective at removing
combinatorial background that tended to accumulate at lower momenta.

\begin{figure}[!hbp]
\centering
\includegraphics[width=\linewidth]{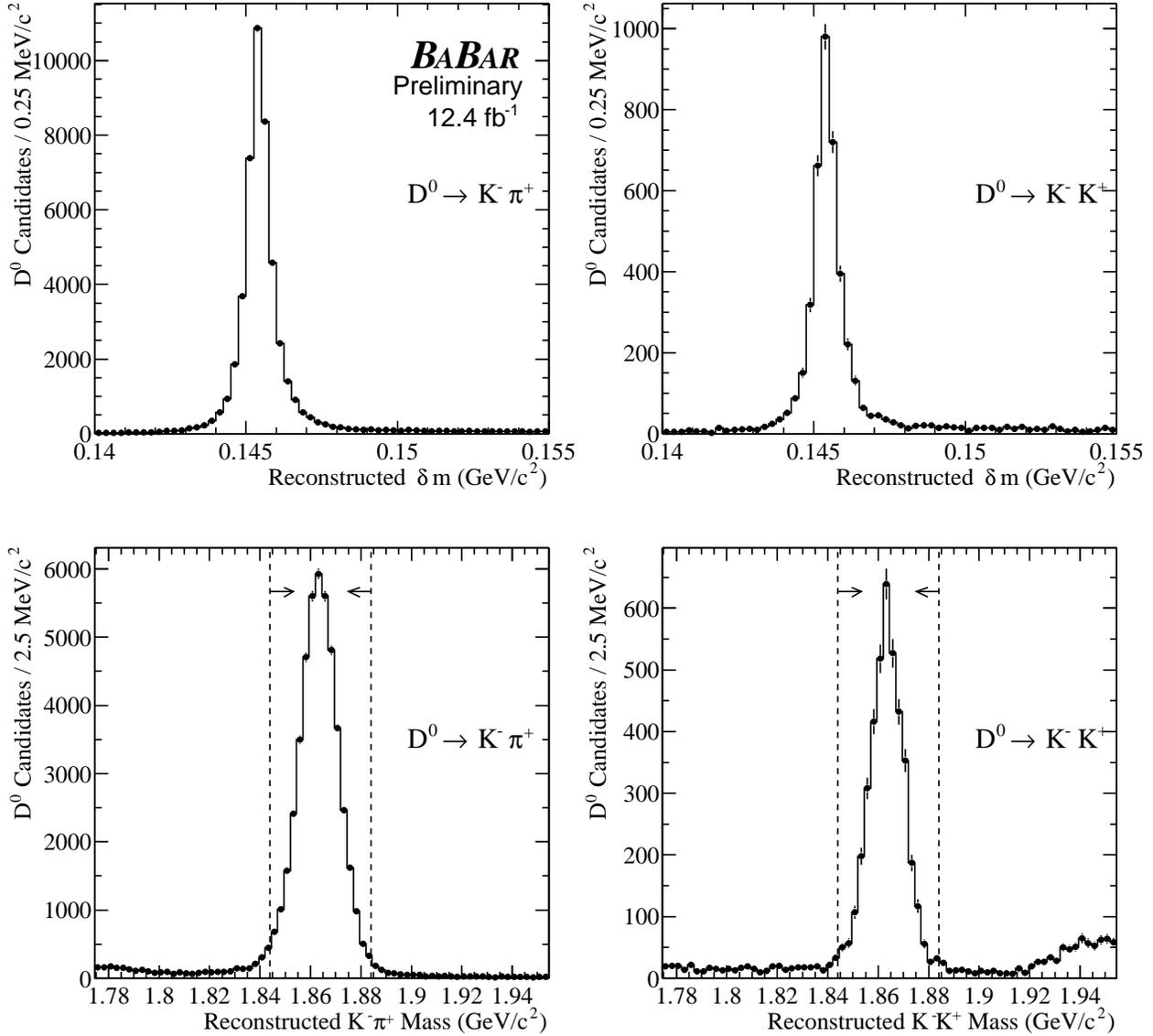}
\caption[This space for rent]{
The reconstructed $\delta m$ and $D^0$ mass distributions
after event selection
for the $K^- \pi^+$ and $K^- K^+$ decay modes. The $\delta m$ plots
include candidates both inside and outside the $\delta m$ selection
requirement but which fall within the $m_D$ window
indicated in the lower plots.
}
\label{fg:dist}
\end{figure}

The $D^0$ mass and
$\delta m$ distribution of the selected events are shown in
Fig.~\ref{fg:dist}. The relative size of the background was about
2\% and 5\% for the $K^-\pi^+$ and $K^-K^+$ samples, respectively,
when measured inside a $\pm 20$\mevcc window. According to
Monte Carlo simulations, of the background in the $K^-\pi^+$ ($K^+ K^-$)
sample, $1/2$ ($1/3$) were combinatorial
background, $1/3$ ($1/6$) were produced by
incorrectly identified $\pi_s$ tracks,
and about $1/6$ ($1/2$) originated from other $D^0$ decays.

\section{Lifetime Determination}

The flight length and its measurement error 
for each $D^0$ candidate were determined by a global,
three dimensional,
multiple vertex fit that included the $D^0$ daughter tracks, 
the $\pi_s$ track,
and the IP envelope. This fit did not include explicit constraints
on the $D^*$ or $D^0$ masses. The value listed by the
Particle Data Group (PDG) for the $D^0$
mass ($m_D = 1.8654$~\gevcc \cite{pdg}) and the momentum of the $D^0$ obtained
with the vertex fit were used to calculate the boost of the $D^0$
and obtain the proper decay time.

An unbinned maximum likelihood fit was used to extract the lifetime
from the $D^0$ samples. The likelihood function was divided into
a decay time distribution function for the signal and a decay time 
distribution function for the background. The signal
function was composed of a convolution of an exponential and
a resolution function. The resolution function was the sum of
two Gaussian distributions with zero mean and with widths that were
proportional to the measurement error (typically 180~fsec)
of the decay time of each $D^0$ candidate.
The parameters in the fit associated with the signal were
the lifetime, the proportional widths of the two Gaussians,
and the fraction of signal that was assigned to the second Gaussian.

Like the signal likelihood function, 
the background function was composed of a convolution of
a resolution function and a lifetime distribution.
The background lifetime distribution was the sum of an
exponential distribution and a delta function at zero, the
latter corresponding to those sources of 
background that originate at the IP.
The resolution function consisted of the sum
of three Gaussian distributions. The first two of these Gaussian
distributions were chosen to match the resolution function of
the signal. The third was given a
width independent of the decay time error and accounted for outliers
produced by long-lived particles or reconstruction errors.
The additional fit parameters associated with the background included the
fraction assigned to zero lifetime sources, 
the background lifetime, the width of the third
Gaussian, and the fraction of background assigned to the third
Gaussian.

\begin{figure}[!hbp]
\centering
\includegraphics[width=\linewidth]{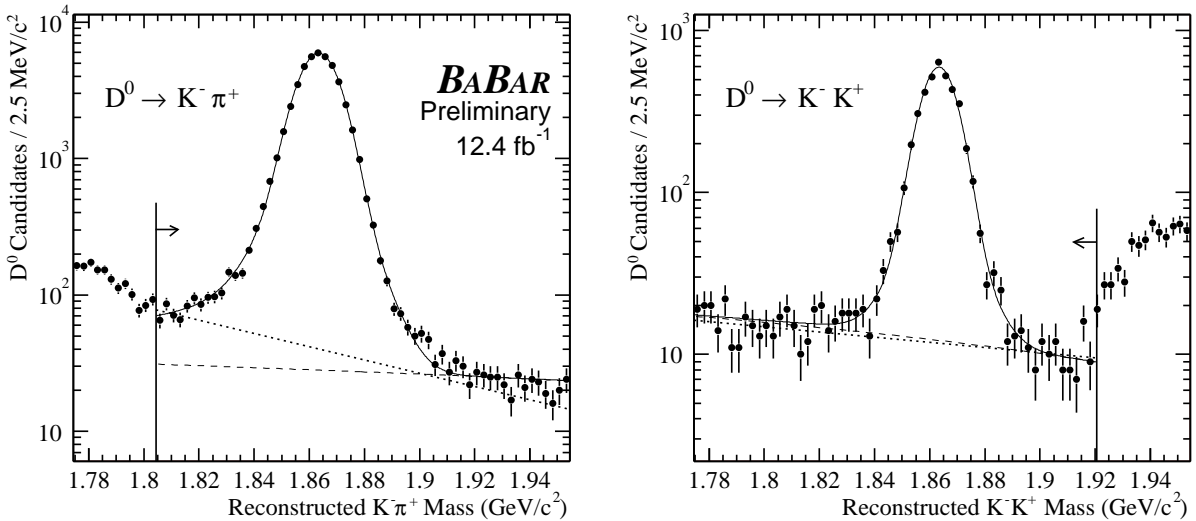}
\caption[This space for rent]{
The fit to the reconstructed $D^0$ mass distribution used to determine
the signal
purity for the lifetime fits. The solid curve is the fit to the
overall distribution and the dashed line is the portion assigned
to the background. The dotted line is an alternative fit of the background
level that is used as a systematic check.
}
\label{fg:mass}
\end{figure}

To combine the signal and background likelihood functions,
the reconstructed mass of each $D^0$ candidate was used to
determine the likelihood that it was part of the signal.
This calculation was based on a separate fit of the
reconstructed $D^0$ mass distribution (Fig.~\ref{fg:mass}).
This fit included a resolution function composed of 
a Gaussian with an asymmetric tail
and a linear portion 
to describe the
background. The slope of the background was constrained with
$D^0$ candidates in the $\delta m$ sideband ($151 < \delta m < 159$\mevcc).

\begin{figure}[!hbp]
\centering
\includegraphics[width=\linewidth]{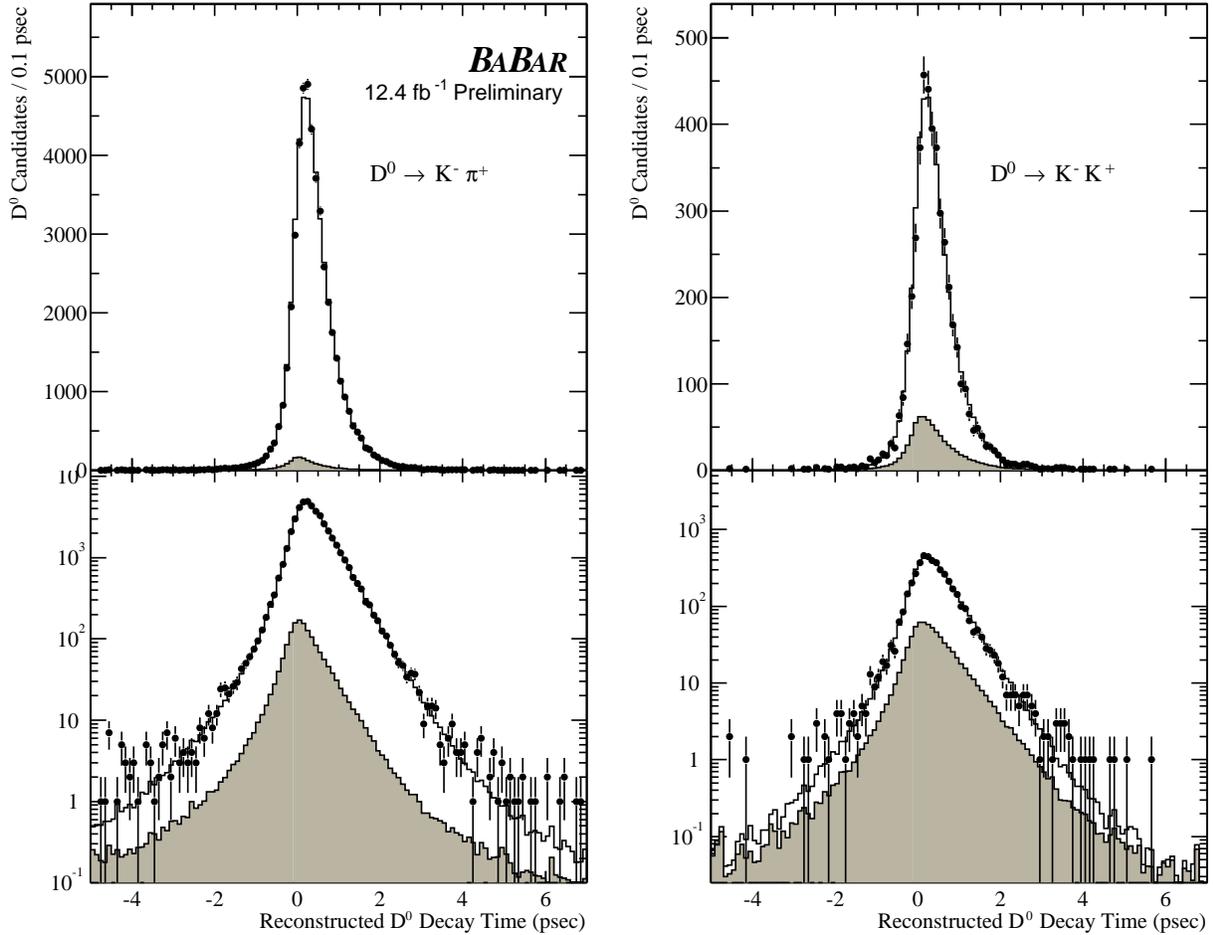}
\caption[This space for rent]{
The fit to the reconstructed $D^0$ lifetime for the two $D^0$ decay modes
for all events including the $D^0$ mass sidebands.
The white histogram represents the result of the unbinned maximum likelihood
fit described in the text. The gray histogram is the portion
assigned to the background.
}
\label{fg:fit}
\end{figure}

The results of the lifetime fits are shown in Fig.~\ref{fg:fit}.
Typical values for the fit parameters were 
a background lifetime similar to the $D^0$ lifetime
and a third Gaussian width that was several times larger than
the typical decay time error. The proportionality factors associated with
the two Gaussians in the resolution function corresponded to
a root-mean-square of approximately $1.2$.

To ensure that the analysis was performed in an objective manner,
the $D^0$ lifetime and $y$ values
were blinded. This blinding was
performed by adding to each of the $\tau_{K\pi}$ and $\tau_{KK}$ fit results
an offset chosen from a random Gaussian distribution
of width 10~fsec. The values of the two offsets and
the positive ($\tau_i > 0$) side of the lifetime
distribution from the data and fit (Fig.~\ref{fg:fit})
were concealed. The value of $y$ was unblinded only
after the analysis method and systematic uncertainties
were finalized and the result was committed for public release.

\section{Systematic Errors and Results}

Many systematic uncertainties cancel
because $y$ was measured from the ratio of lifetimes.
The few uncertainties that do not
were associated mostly with backgrounds. These were tested by varying
each event selection requirement within its uncertainty and 
recording the subsequent change $\Delta_i$ in the measured value
of $y$.
The quadrature difference
$(\delta \Delta_i)^2 = | \sigma_0^2 - \sigma_i^2 |$
was used as an estimate of the
statistical error $\delta \Delta_i$ in $\Delta_i$,
where $\sigma_0$ ($\sigma_i$)
was the statistical error in $y$ before (after)
the $i{\hskip 0.5pt}$th systematic
check. Each systematic check with $\Delta_i > \delta\Delta_i$
was included in the sum $\sum \Delta_i^2 - (\delta\Delta_i)^2$.
The square root of this sum (1.7\%) was used as the estimate for the
systematic uncertainty from event selection and
background.

Biases in tracking reconstruction
were explored by
studying Monte Carlo samples, which, within statistics,
showed no reconstruction bias. 
In addition, the lifetimes were 
compared to measurements that employ
a variety of vertexing techniques, including constraining the
$D^0$ mass and using separate $D^*$ and $D^0$ vertex fits.
A systematic uncertainty of 0.4\% was assigned as a result.

Detector misalignment was another potential source of bias.
Systematic distortions of the SVT,
even as small as a few microns, can produce significant
variations in the apparent $D^0$ lifetime. Several studies were
used to measure and characterize such distortions, and
strategies were developed to correct them. 
One example was the study of
proton tracks that were created by the interaction of off-energy
beam particles and the beampipe. These tracks were used to
measure the radius of the beampipe to a precision of a few microns,
which limited the uncertainty in the radial scale of
the SVT to three parts in one thousand.

Another example was a study of
$e^+ e^- \rightarrow \gamma\gamma \rightarrow 4\pi^\pm$ events
in which the four charged tracks were known to originate from the IP.
By selecting oppositely charged pairs of these tracks with opening
angles similar to two-prong $D^0$ decays, it was possible to
measure the apparent IP position as a function of $D^0$ trajectory
and calculate a correction to the $D^0$ lifetime. For the data
sample used in this analysis, this correction was determined
to be $+5$~fsec, with negligible statistical error and a systematic
uncertainty of $\pm5$~fsec. This type of correction nearly
cancels in the lifetime ratio and introduces little systematic uncertainty
in $y$.

\begin{table}[!hpb]
\caption[This space for rent]{
Individual contributions to the systematic uncertainty in $y$.
}
\label{tb:syst}
\begin{center}
\begin{tabular}{lc}
\hline
\hline
Category & Uncertainty (\%)         \\
\hline
Event Selection and Background      & $1.7$ \\
Reconstruction and Vertexing        & $0.4$ \\
Alignment                    	    & $0.3$ \\
\hline
Quadrature Sum                      & $1.7$ \\
\hline
\hline
\end{tabular}
\end{center}
\end{table}

The systematic uncertainties in $y$ 
are summarized in Table~\ref{tb:syst}.
When all systematic checks are added in quadrature, 
the preliminary result for $y$ is:
\begin{equation*}
y = \left( -1.0 \pm 2.2 \pm 1.7 \right)\;\% \;,
\end{equation*}
where the first error is statistical and the second, systematic.

The same set of systematic checks was applied to the $D^0$ lifetime.
In this case, several of the systematic uncertainties, in particular
those corresponding to detector alignment, do not cancel as they do
for the lifetime ratio, and as a result, the total systematic
uncertainty was dominated by different sources.
Nevertheless, an important test of the $y$ analysis
was a $D^0$ lifetime that agreed with expectations.
A corrected value of $\tau_{K\pi} = 412 \pm 2$~fsec was found with
a systematic uncertainty of approximately 6~fsec, which is
consistent with the PDG value of $412.6 \pm 2.8$~fsec \cite{pdg}.

\section{Conclusion}

The following preliminary value of $y$ was measured
using 12.4\invfb of data taken by the \babar\  detector
in 2001:
\begin{equation*}
y = \left( -1.0 \pm 2.2 \pm 1.7 \right)\;\% \;,
\end{equation*}
where the first error is statistical and the second, systematic.
This result is consistent with the Standard Model expectation of zero
and consistent with published values from
E791 \cite{e791} and FOCUS \cite{focus}
and preliminary results from Belle \cite{belle} and CLEO \cite{cleo}.

The measurement reported in this paper is currently limited by statistics.
As additional data are collected and as previous data are reprocessed
with the latest alignment parameters, the statistical uncertainty is expected
to decrease significantly.

\section{Acknowledgements}

We are grateful for the 
extraordinary contributions of our \pep2\ colleagues in
achieving the excellent luminosity and machine conditions
that have made this work possible.
The collaborating institutions wish to thank 
SLAC for its support and the kind hospitality extended to them. 
This work is supported by the
US Department of Energy
and National Science Foundation, the
Natural Sciences and Engineering Research Council (Canada),
Institute of High Energy Physics (China), the
Commissariat \`a l'Energie Atomique and
Institut National de Physique Nucl\'eaire et de Physique des Particules
(France), the
Bundesministerium f\"ur Bildung und Forschung
(Germany), the
Istituto Nazionale di Fisica Nucleare (Italy),
the Research Council of Norway, the
Ministry of Science and Technology of the Russian Federation, and the
Particle Physics and Astronomy Research Council (United Kingdom). 
Individuals have received support from the Swiss 
National Science Foundation, the A. P. Sloan Foundation, 
the Research Corporation, the Della Riccia Foundation,
and the Alexander von Humboldt Foundation.


\begin{thebibliography}{99}

\bibitem{theory} F.~Buccella, M.~Lusignoli, and A.~Publiese,
\plb{379} (1996) 249; F.~Buccella {\it et al.}, \jprd{51} (1995) 3478.

\bibitem{phenom} E.~Golowich and S.~Pakvasa, \plb{505} (2001) 94.

\bibitem{pdg}  D.E.~Groom {\it et al.}, \epjc{15} (2000) 1.

\bibitem{babar}
The \babar\ Collaboration, B.~Aubert {\it et al.},
to appear in Nucl.\ Instrum.\ Methods, \linebreak hep--ex/0105044.

\bibitem{e791} The E791 Collaboration, E.M.~Aitala {\it et al.},
\jprl{83} (1999) 32.

\bibitem{focus} The Focus Collaboration, 
J.M.~Link {\it et al.}, \plb{485} (2000) 62.

\bibitem{belle} The Belle Collaboration, 
talk presented by B.D.~Yabsley at the 
International Europhysics Conference on High
Energy Physics (EPS HEP 2001), Budapest (2001) BELLE--CONF--0131.

\bibitem{cleo} The CLEO Collaboration, 
to appear in the proceedings of the 4th International Conference
on $B$ Physics and \CP Violation (BCP 4), Japan (2001) hep--ex/0104008.

\end{thebibliography}
\end{document}